\documentclass[10pt,a4paper]{article}
\usepackage{graphicx}
\usepackage{multicol}
\usepackage{tabto}
\usepackage{url}
\usepackage{enumitem}
\usepackage{textcomp}
\usepackage{eurosym}
\usepackage{xcolor}
\usepackage[labelfont=bf]{caption}
\usepackage[yyyymmdd]{datetime}

\usepackage{flowchart}
\usetikzlibrary{shapes,arrows.meta,chains,external}

\newcommand{\ts}{
    \tikzset{
        >={Latex[width=3mm,length=3mm]},
        x = 3em,
        y = 3em
    }
    \tikzstyle{line} = [draw, ->, >=latex, thick]
    \tikzstyle{circ} = [
        circle,
        draw,
        thick,
        align=center,
        minimum width=0.3em,
        minimum height=0.3em,
        fill=magenta!60
    ]
    \tikzstyle{block} = [
        rectangle,
        draw,
        thick,
        text centered,
        rounded corners,
        align=center,
        text width=\smbwd,
        minimum width=\smbwd,
        minimum height=1em,
        fill=orange!30
    ]
    \tikzstyle{noshape} = [text width=5em, text centered, minimum height=5em]
    \tikzstyle{nos} = [
        rectangle,
        text width=6em,
        text centered,
        align=center,
        minimum height=1cm
    ]
    \tikzstyle{box} = [
      draw,
      rounded corners,
      rectangle,
      align=left,
      text width=18em,
      minimum width=1cm,
      minimum height=1cm
    ]
}

\begin{document}

\bibliographystyle{plain}
\thispagestyle{empty}

\begin{center}
\large{\bf Serverless Electronic Mail}\\
\end{center}
\vspace{0.5em}
\begin{center}
{\large Geoffrey Goodell}\\
\vspace{0.5em}
{\large University College London}\\
\vspace{0.5em}
{\large \texttt{g.goodell@ucl.ac.uk}}
\end{center}

\begin{center}
{\textit{This Version: \today}}\\
\end{center}
\vspace{4em}

\begin{abstract}

We describe a simple approach to peer-to-peer electronic mail that would allow
users of ordinary workstations and mobile devices to exchange messages without
relying upon third-party mail server operators.  Crucially, the system allows
participants to establish and use multiple unlinked identities for
communication with each other.  The architecture leverages ordinary
SMTP~\cite{smtp} for message delivery and Tor~\cite{tor} for peer-to-peer
communication.  The design offers a robust, unintrusive method to use
self-certifying Tor onion service names to bootstrap a web of trust based on
public keys for end-to-end authentication and encryption, which in turn can be
used to facilitate message delivery when the sender and recipient are not
online simultaneously.  We show how the system can interoperate with existing
email systems and paradigms, allowing users to hold messages that others can
retrieve via IMAP~\cite{imap} or to operate as a relay between system
participants and external email users.  Finally, we show how it is possible to
use a gossip protocol to implement mailing lists and how distributed ledger
technology might be used to bootstrap consensus about shared knowledge among
list members.

\end{abstract}

\section{Objectives}

Third-party e-mail platforms operate control points that can, and often do,
function against the interests and purposes of their users and the other e-mail
users with whom their users correspond.  However, the convenience offered by
such platforms, coupled with the architecture of carrier networks and the
mobility of user devices, typically prevents individual users from operating
their own mail servers.  In particular, many broadband and mobile carriers
implement firewalls, either explicitly or \textit{de facto} via network address
translation, to block packets destined for servers operated by their clients.
Carriers without such policies often assign their users dynamic addresses,
rendering them unreachable as a means of receiving mail and untrusted by remote
servers as a means of sending mail.  As a result, although server software can
certainly run on mobile devices such as laptops and phones, few users actually
run such software, and platform operators such as Apple and Google have little
incentive to promote a departure from this paradigm.

Tor onion services~\cite{tor-onion} offer an accepted way for users to reach
each other directly.  The operator of an onion service reaches out through the
Tor network to establish a point of presence that other users can access over
an end-to-end encrypted channel using a self-certifying name.  Although the Tor
relays, including the introduction and rendezvous points, facilitate the
end-to-end connection between the Tor client and the onion service, they have
no knowledge of the identities of the clients and services that they are
connecting.  Tor software runs on most mobile devices, and mail servers can be
compiled to run on most mobile devices as well.  It is therefore entirely
possible for mobile devices such as phones and tablets, in addition to ordinary
laptops and workstations, to run a combination of software utilities that can
underpin a peer-to-peer e-mail network, and a package containing such utilities
can be designed to interoperate with popular mail clients with minimal
modification.

Next, we define the specific goals of our project, which we believe will
increase the value of e-mail to its users and promote trust in the
infrastructure that delivers it:

\begin{enumerate}

\item \textit{Require end-to-end encryption between users.}  Although popular
mail servers tout their use of client-to-server encryption as well as
server-to-server authentication via SPF~\cite{spf} and DKIM~\cite{dkim},
ordinary customers of popular e-mail platforms still do not generally use
end-to-end encryption technology.  In contrast, our system would leverage
peer-to-peer trust to provide real, usable security for its users.

\item \textit{Eliminate third-party mail server operators.}  Electronic mail
generally relies on network carriers to deliver messages from the source to the
destination.  If the sender and recipient are online simultaneously, then they
can leverage anonymity networks to communicate without third-party mail servers
to collect metadata about their conversations.  Even when the sender and
recipient are not online at the same time, there is no reason that these
carriers must be giant platforms.  In contrast, our system would leverage
peer-to-peer relationships to deliver messages.

\item \textit{Support multiple unlinkable identities per user.}  Most people
use a small number of distinct email addresses to receive mail from a large
number of different users.  As a result, they implicitly create linkages among
their many relationships that can be observed by third parties.  Our system
natively supports the establishment of an arbitrary number of unlinkable
identities for each user.  By not providing a way for an individual to prove
that two identities are linked, we reduce the chance that linkages will be
forcibly discovered by an adversary.

\end{enumerate}

\section{Comparison to Other Projects}

Anonymous, web-based e-mail accounts are already available on the Internet.
Unfortunately, such e-mail accounts are generally incompatible with ordinary
mail client software and business processes, and they usually have few features
and service-level assurances.  They also rely upon third-party operators, who
might block access or simply stop working at any time, thus resisting the
establishment of long-term identities or addresses.  So we need something
different.  There are a few notable projects that put the users in the center
of the architecture; we compare our system to theirs in terms of requirements:

\begin{enumerate}

\item \textit{Mixminion,} a ``Type III Anonymous Remailer''~\cite{mixminion}.
Mixminion uses mix networks to batch, mix, and resend anonymous e-mail messages
through a network of remailers.  Mixminion extends the
Mixmaster~\cite{mixmaster} protocol in several ways, notably by introducing
reply blocks to facilitate secure replies to an anonymous sender.  Mixminion is
designed to tolerate high latency, so theoretically its anonymity properties
could be better than Tor, which is designed for low-latency applications and
therefore introduces a vulnerability to timing attacks.  Although Mixminion has
been released, it has not been under active development since 2013, and its
developers do not recommend its use~\cite{mixminion-web}.  In contrast, our
system is explicitly designed for low-latency operation, in part because we
believe users do not want to wait for their mail, and in part because the Tor
anonymity network delivers a much larger anonymity set that has already been
bootstrapped.  Additionally, although our system can be used to send and
receive anonymous e-mail messages, we do not anticipate that will be its
typical use case.  We imagine that in most circumstances the senders and
recipients will know who each other are.  Nevertheless, the system relies upon
the anonymity of the Tor network to protect users from network adversaries and
to ensure that the application itself is not blocked.

\item \textit{Cwtch,} an ``Infrastructure for Asynchronous, Decentralized,
Multi-Party, and Metadata Resistant Applications''~\cite{cwtch}.  Cwtch is an
extension of the Ricochet~\cite{ricochet} protocol, which provides real-time
instant messaging using Tor onion services.  Cwtch extends Ricochet by allowing
asynchronous group messaging.  In contrast, despite the fact that our proposal
anticipates low latency, our system is not principally intended for real-time
instant messaging.  We assume that typical messages can be lengthy, with rich
features and attached files, and we assume that typical users will want to read
each one individually and at their convenience.  We further assume that users
will often be offline when communication takes place, as they would for
ordinary e-mail.  For all of these reasons, our system does not use Ricochet
and instead relies upon ordinary SMTP for passing messages.

\end{enumerate}

\section{System Requirements}

We imagine that a user of our proposed system will have access to the Internet
and a device that can accommodate Tor client software, an SMTP client, and an
SMTP server.  Nearly all modern workstations and laptops running Linux,
Windows, MacOS, or a BSD derivative will satisfy the device requirement.
Smartphones and tablets will usually satisfy this requirement as well, although
users are advised that, depending on the specific restrictions imposed by their
device vendors, they might need to install a custom operating system to install
the requisite software applications.\footnote{Important differences exist
between the iOS and Android development platforms.  Although data-harvesting
might be a larger part of the revenue model for Google, Apple's role as
gatekeeper in the iOS platform might encumber an effort to deploy systems like
ours~\cite{greene2017}.}

The Tor client software must be configured to allow a helper application to
operate persistent Tor onion services via the Tor Control
Protocol~\cite{tor-control}.  The user shall also have compliant e-mail
software, including a modified mail server that can accommodate the behaviour
defined in this document.  The user should also have a modified mail client
that can facilitate the behaviour defined in this document.  Depending upon how
the server is implemented, such modifications might not be strictly required,
although we imagine that appropriate client modifications could significantly
improve the user experience and value of this system.  The user should also
have suitable PGP software~\cite{zimmermann1991}; we assume this will be
OpenPGP~\cite{gpg}.

\section{System Design Overview}

\begin{figure}
\begin{center}
\begin{tikzpicture}[
    >=latex,
    font=\sf,
    auto
]\ts
\def\smbwd{8em}

\node (r1) at (0,0) [noshape, text width=4em] {
    \scalebox{0.1}{\includegraphics{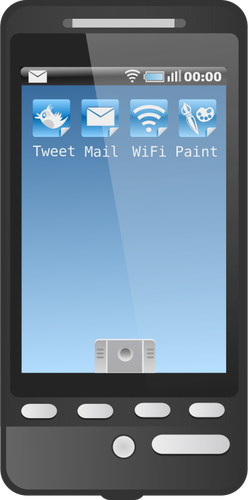}}
};
\node (r2) at (5,0) [noshape, text width=4em] {
    \scalebox{0.1}{\includegraphics{images/1281043443.png}}
};

\draw[->, line width=0.5mm] (r1) -- node[above] {
    \scalebox{0.1}{\includegraphics{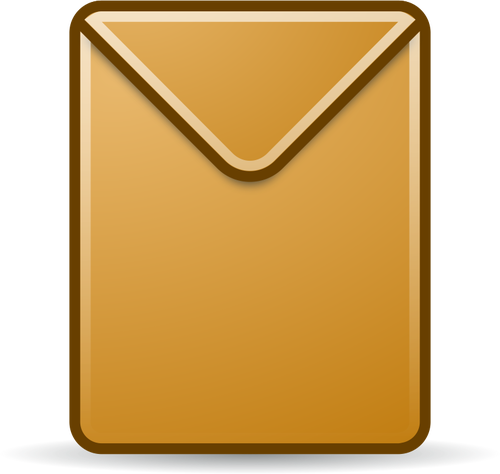}}
} node[below] {
    \scalebox{0.2}{\includegraphics{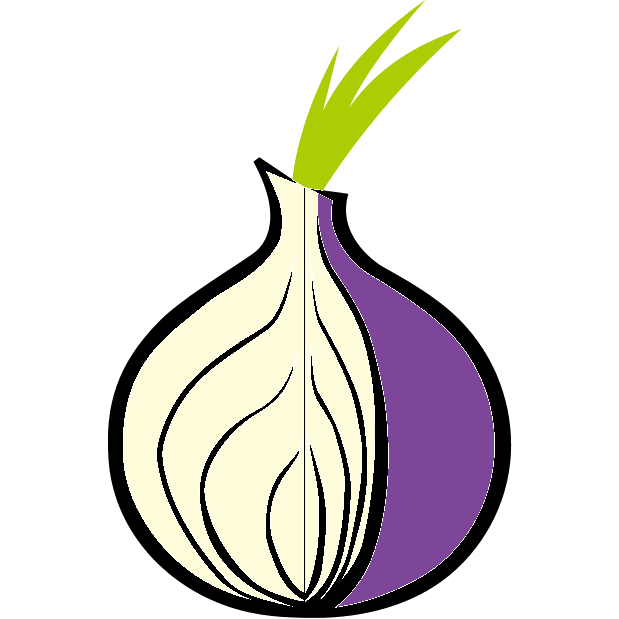}}
} (r2);

\node (m1) at (2.5, 3.4) [
    box,
    yshift=-5pt,
    color=black,
    fill=orange!30
] {
    \small{From: Alice \textless{user@wxu6pped7wv3.onion}\textgreater}\\
    \small{To: Bob \textless{user@0xiq82snnr09.onion}\textgreater}\\
    \vspace{1.25em}
    \small{Message: ...}
};
\draw[->, line width=0.5mm] (m1) -- (2.5, 1.8);

\node (d1) at (0,-1.2) [nos] {Alice};
\node (d1) at (5,-1.2) [nos] {Bob};

\end{tikzpicture}

\caption{\textit{If Alice runs a mail client and server on her device and Bob
does the same, then Alice can send mail directly to Bob using Tor onion
services.  The channel from Alice's Tor client to Bob's Tor client is protected
by end-to-end encryption. }}

\label{f:ab}
\end{center}
\end{figure}

The elemental feature of our design is pairwise communication between two
parties via SMTP over Tor, wherein each user operates at least one (and
probably more than one) Tor onion service that relays traffic to an SMTP
server, as illustrated by Figure~\ref{f:ab}.  The SMTP servers are not
generally intended to be available outside the Tor network, and that therefore
the system does not rely upon Tor exit relays for its operation.  Most Tor exit
relays disallow traffic exiting to common TCP ports used for SMTP anyway,
although this is not a concern for the system we describe.

\subsection{Personal Introductions}

\begin{figure}
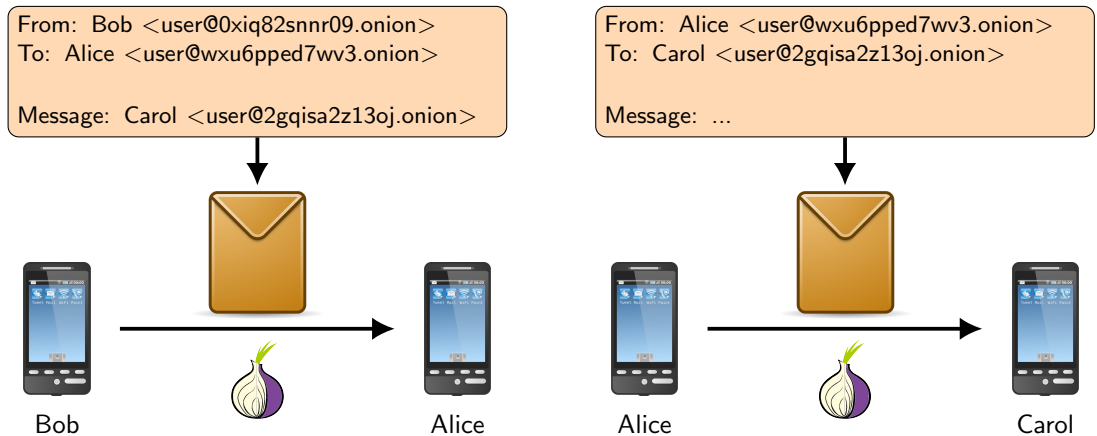

\begin{center}
\begin{tikzpicture}[
    >=latex,
    font=\sf,
    auto
]\ts
\def\smbwd{8em}

\node (r1) at (0,0) [noshape, text width=4em] {
    \scalebox{0.1}{\includegraphics{images/1281043443.png}}
};
\node (r2) at (5,0) [noshape, text width=4em] {
    \scalebox{0.1}{\includegraphics{images/1281043443.png}}
};

\draw[->, line width=0.5mm] (r1) -- node[above] {
    \scalebox{0.1}{\includegraphics{images/rodentia-icons_package-x-generic.png}}
} node[below] {
    \scalebox{0.2}{\includegraphics{images/onion.png}}
} (r2);

\node (m1) at (2.5, 3.4) [
    box,
    yshift=-5pt,
    color=black,
    fill=orange!30
] {
    \small{From: Bob \textless{user@0xiq82snnr09.onion}\textgreater}\\
    \small{To: Alice \textless{user@wxu6pped7wv3.onion}\textgreater}\\
    \vspace{1.25em}
    \small{Message: Carol \textless{user@2gqisa2z13oj.onion}\textgreater}
};
\draw[->, line width=0.5mm] (m1) -- (2.5, 1.8);

\node (d1) at (0,-1.2) [nos] {Bob};
\node (d1) at (5,-1.2) [nos] {Alice};

\end{tikzpicture}
\begin{tikzpicture}[
    >=latex,
    font=\sf,
    auto
]\ts
\def\smbwd{8em}

\node (r1) at (0,0) [noshape, text width=4em] {
    \scalebox{0.1}{\includegraphics{images/1281043443.png}}
};
\node (r2) at (5,0) [noshape, text width=4em] {
    \scalebox{0.1}{\includegraphics{images/1281043443.png}}
};

\draw[->, line width=0.5mm] (r1) -- node[above] {
    \scalebox{0.1}{\includegraphics{images/rodentia-icons_package-x-generic.png}}
} node[below] {
    \scalebox{0.2}{\includegraphics{images/onion.png}}
} (r2);

\node (m1) at (2.5, 3.4) [
    box,
    yshift=-5pt,
    color=black,
    fill=orange!30
] {
    \small{From: Alice \textless{user@wxu6pped7wv3.onion}\textgreater}\\
    \small{To: Carol \textless{user@2gqisa2z13oj.onion}\textgreater}\\
    \vspace{1.25em}
    \small{Message: ...}
};
\draw[->, line width=0.5mm] (m1) -- (2.5, 1.8);

\node (d1) at (0,-1.2) [nos] {Alice};
\node (d1) at (5,-1.2) [nos] {Carol};

\end{tikzpicture}

\caption{\textit{Bob can introduce Alice to Carol by including contact details
for Carol in a message to Alice.  Then Alice can send a message to Carol
directly.  The combination of Bob's introduction and the self-certifying Onion
address form the basis for trusting Carol.}}

\label{f:bac}
\end{center}
\end{figure}

Before they can send mail to each other via our system, the users will need to
have received the in-system e-mail addresses of their counterparties.  This can
be accomplished in several ways:

\begin{enumerate}

\item \textit{Manually added by the user.}

\begin{enumerate}

\item \textit{One-to-one messaging via an external channel.} Of course, it is
possible for a community to share such addresses with each other by exchanging
files over any other digital medium, including regular messaging with servers
to support e-mail or chat.

\item \textit{Via a web page or message board.} Web pages or message boards
could advertise e-mail addresses for use within this system.

\item \textit{In person}. This could be done with near-field communication or
QR-code scanning by a mobile device.  The QR-code scanning could be
device-to-device, or it could involve a device scanning a posted or printed
advertisement.

\end{enumerate}

\item \textit{Automatically added via communication within the system.}

\begin{enumerate}

\item \textit{In-system introductions.}  Once two counterparties know how to
reach each other, they can introduce each other to third parties, as
illustrated in Figure~\ref{f:bac}.

\item \textit{Mailing lists.}  One user can invite another to participate in a
mailing list by sharing the name of the list and agreeing to forward list
messages in one or both directions.  For example, Bob can agree to forward
messages from Alice to the list, or from the list to Alice, or both; we
describe mailing lists in Section~\ref{ss:lists}.

\end{enumerate}

\end{enumerate}

\subsection{Message Delivery}

\begin{figure}
\begin{center}
\begin{tikzpicture}[
    >=latex,
    font=\sf,
    auto
]\ts
\def\smbwd{8em}

\node (r1) at (0,0) [noshape, text width=4em] {
    \scalebox{0.1}{\includegraphics{images/1281043443.png}}
};
\node (r2) at (5,0) [noshape, text width=4em] {
    \scalebox{0.1}{\includegraphics{images/1281043443.png}}
};

\draw[->, line width=0.5mm] (r1) edge[bend left=25] node[above] {
} node[below] {
    \scalebox{0.2}{\includegraphics{images/onion.png}}
} (r2);
\draw[->, line width=0.5mm] (r2) edge[bend left=25] node[above] {} (r1);

\node (m1) at (2.5, 2.4) [
    box,
    yshift=-5pt,
    color=black,
    fill=orange!30
] {
    \small{From: Alice \textless{user@wxu6pped7wv3.onion}\textgreater}\\
    \small{To: Carol \textless{user@2gqisa2z13oj.onion}\textgreater}\\
    \small{OpenPGP Key Request}
};
\draw[->, line width=0.5mm] (m1) -- (2.5, 0.8);

\node (m2) at (2.5, -2) [
    box,
    yshift=-5pt,
    color=black,
    fill=orange!30
] {
    \small{From: Carol \textless{user@2gqisa2z13oj.onion}\textgreater}\\
    \small{To: Alice \textless{user@wxu6pped7wv3.onion}\textgreater}\\
    \small{OpenPGP Key Response: \textbf{0k4jnr1l701a}}
};
\draw[->, line width=0.5mm] (m2) -- (2.5, -0.8);

\node (d1) at (0,-1.2) [nos] {Alice};
\node (d1) at (5,-1.2) [nos] {Carol};

\end{tikzpicture}
\begin{tikzpicture}[
    >=latex,
    font=\sf,
    auto
]\ts
\def\smbwd{8em}

\node (r1) at (0,0) [noshape, text width=4em] {
    \scalebox{0.1}{\includegraphics{images/1281043443.png}}
};
\node (r2) at (5,0) [noshape, text width=4em] {
    \scalebox{0.1}{\includegraphics{images/1281043443.png}}
};

\draw[->, line width=0.5mm] (r1) -- node[above] {
    \scalebox{0.1}{\includegraphics{images/rodentia-icons_package-x-generic.png}}
} node[below] {
    \scalebox{0.2}{\includegraphics{images/onion.png}}
} (r2);
\node (i1) at (2.5, 0.95) {
    \scalebox{0.07}{\includegraphics{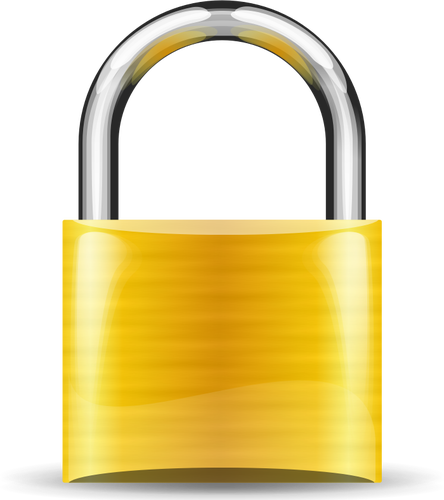}}
};

\node (m1) at (2.5, 3.4) [
    box,
    yshift=-5pt,
    color=black,
    fill=orange!30
] {
    \small{From: Alice \textless{user@wxu6pped7wv3.onion}\textgreater}\\
    \small{To: Carol \textless{user@2gqisa2z13oj.onion}\textgreater}\\
    \small{For: Carol \textless{user@2gqisa2z13oj.onion}\textgreater}\\
    \small{Message: [...] encrypted to \textbf{0k4jnr1l701a}}
};
\draw[->, line width=0.5mm] (m1) -- (2.5, 1.8);

\node (d1) at (0,-1.2) [nos] {Alice};
\node (d2) at (5,-1.2) [nos] {Carol};
\node (d3) at (2.5,0.7) [nos] {\Large{\textbf{E}}};

\end{tikzpicture}

\caption{\textit{Once Alice knows how to reach Carol, she can request Carol's
OpenPGP key.  This step can be implemented automatically via the headers of
messages that Alice and Carol send to each other, for example using reply
blocks.  With Carol's OpenPGP key, Alice can send encrypted messages to Carol that
can be stored and forwarded if required.  (Note: the padlock with the `E'
symbol on the envelope indicates that the message is encrypted from Alice's
message-writing application to Carol's message-reading application.)}}

\label{f:ac}
\end{center}
\end{figure}

Once two parties are able to communicate with each other using our protocol,
then they can rely upon Tor to ensure that the channel over the network between
them is encrypted.  However, the use of Tor onion services does not
automatically ensure that their applications can verify the authenticity of
messages, and indeed there is no specific protection for the message between
the endpoint of the Tor service and the recipient's mail client.  This is
important for two reasons.  First, most mail clients assume that mail is
persistent, not ephemeral, and OpenPGP~\cite{gpg}, the established mechanism
for exchanging end-to-end encrypted messages, would operate outside the
security envelope of Tor.  Second, users might want third-parties to carry
messages for them, a desideratum that is particularly important when the
recipient is not online at the time that the sender sends the message.

To address this concern, we specify a mechanism by which two parties can
mutually exchange their OpenPGP keys~\cite{gpg}, relying upon the security
envelope provided by Tor to ensure that the message is not intercepted by
network adversaries.  Once the users have exchanged their OpenPGP keys in this
manner, they can exchange messages that are end-to-end encrypted and suitable
for persistent storage, as shown in Figure~\ref{f:ac}.

\begin{figure}
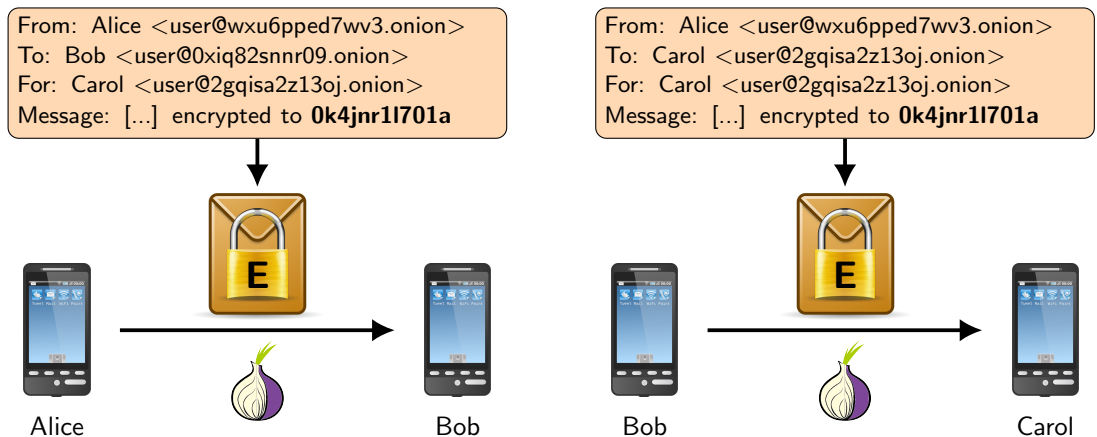

\begin{center}
\begin{tikzpicture}[
    >=latex,
    font=\sf,
    auto
]\ts
\def\smbwd{8em}

\node (r1) at (0,0) [noshape, text width=4em] {
    \scalebox{0.1}{\includegraphics{images/1281043443.png}}
};
\node (r2) at (5,0) [noshape, text width=4em] {
    \scalebox{0.1}{\includegraphics{images/1281043443.png}}
};

\draw[->, line width=0.5mm] (r1) -- node[above] {
    \scalebox{0.1}{\includegraphics{images/rodentia-icons_package-x-generic.png}}
} node[below] {
    \scalebox{0.2}{\includegraphics{images/onion.png}}
} (r2);
\node (i1) at (2.5, 0.95) {
    \scalebox{0.07}{\includegraphics{images/Padlock-gold.png}}
};

\node (m1) at (2.5, 3.4) [
    box,
    yshift=-5pt,
    color=black,
    fill=orange!30
] {
    \small{From: Alice \textless{user@wxu6pped7wv3.onion}\textgreater}\\
    \small{To: Bob \textless{user@0xiq82snnr09.onion}\textgreater}\\
    \small{For: Carol \textless{user@2gqisa2z13oj.onion}\textgreater}\\
    \small{Message: [...] encrypted to \textbf{0k4jnr1l701a}}
};
\draw[->, line width=0.5mm] (m1) -- (2.5, 1.8);

\node (d1) at (0,-1.2) [nos] {Alice};
\node (d2) at (5,-1.2) [nos] {Bob};
\node (d3) at (2.5,0.7) [nos] {\Large{\textbf{E}}};

\end{tikzpicture}
\begin{tikzpicture}[
    >=latex,
    font=\sf,
    auto
]\ts
\def\smbwd{8em}

\node (r1) at (0,0) [noshape, text width=4em] {
    \scalebox{0.1}{\includegraphics{images/1281043443.png}}
};
\node (r2) at (5,0) [noshape, text width=4em] {
    \scalebox{0.1}{\includegraphics{images/1281043443.png}}
};

\draw[->, line width=0.5mm] (r1) -- node[above] {
    \scalebox{0.1}{\includegraphics{images/rodentia-icons_package-x-generic.png}}
} node[below] {
    \scalebox{0.2}{\includegraphics{images/onion.png}}
} (r2);
\node (i1) at (2.5, 0.95) {
    \scalebox{0.07}{\includegraphics{images/Padlock-gold.png}}
};

\node (m1) at (2.5, 3.4) [
    box,
    yshift=-5pt,
    color=black,
    fill=orange!30
] {
    \small{From: Alice \textless{user@wxu6pped7wv3.onion}\textgreater}\\
    \small{To: Carol \textless{user@2gqisa2z13oj.onion}\textgreater}\\
    \small{For: Carol \textless{user@2gqisa2z13oj.onion}\textgreater}\\
    \small{Message: [...] encrypted to \textbf{0k4jnr1l701a}}
};
\draw[->, line width=0.5mm] (m1) -- (2.5, 1.8);

\node (d1) at (0,-1.2) [nos] {Bob};
\node (d2) at (5,-1.2) [nos] {Carol};
\node (d3) at (2.5,0.7) [nos] {\Large{\textbf{E}}};

\end{tikzpicture}

\caption{\textsc{Message Carriers.} \textit{If Carol is not online when Alice
tries to send a message, then Alice can encrypt the message with Carol's OpenPGP
key and send it to Bob, whose software can subsequently try to send the message
to Carol on behalf of Alice, even if Alice goes offline.  Note that SMTP
explicitly recommends a way for Bob to periodically resend the message until it
succeeds.}}

\label{f:abc}
\end{center}
\end{figure}

\begin{figure}
\begin{center}
\begin{tikzpicture}[
    >=latex,
    font=\sf,
    auto
]\ts
\def\smbwd{8em}

\node (r1) at (0,0) [noshape, text width=4em] {
    \scalebox{0.1}{\includegraphics{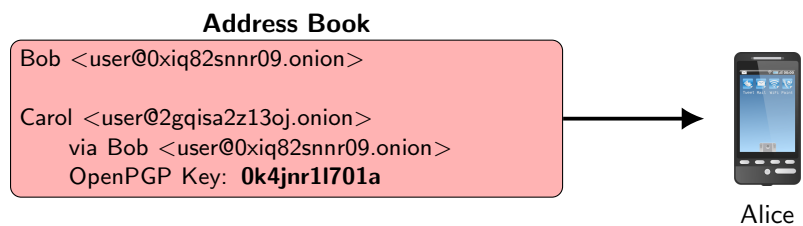}}
};

\node (m1) at (-6, 0) [
    box,
    color=black,
    text width=20em,
    fill=red!30
] {
    \small{Bob \textless{user@0xiq82snnr09.onion}\textgreater}\\
    \vspace{1.25em}
    \small{Carol \textless{user@2gqisa2z13oj.onion}\textgreater}\\
    \small{\hspace{2em}via Bob \textless{user@0xiq82snnr09.onion}\textgreater}\\
    \small{\hspace{2em}OpenPGP Key: \textbf{0k4jnr1l701a}}\\
};
\draw[->, line width=0.5mm] (m1) -- (r1);

\node (d1) at (0,-1.2) [nos] {Alice};
\node (d2) at (-6,1.2) [nos, text width=10em] {\textbf{Address Book}};

\end{tikzpicture}

\caption{\textsc{Address Book Function.} \textit{It is assumed that Alice's
mail client will maintain a mapping that includes metadata such as the
provenance of the introduction, designated carriers of messages, and public
cryptographic keys for each contact.}}

\label{f:book}
\end{center}
\end{figure}

Users who are able to send end-to-end encrypted messages can also rely upon
\textit{carriers} to forward messages on their behalf, as shown in
Figure~\ref{f:abc}.  This feature is useful when the recipient is not online at
the time that a message is sent; in such circumstances we propose that a third
party can relay the message at the behest of the sender, thus addressing the
availability problem that mail servers intend to solve.  Carriers using SMTP
will be able to perform best-effort delivery just as an ordinary mail server
would.  The main benefit of using a carrier is to decouple the timeliness of
delivery from the sender's online status, thus improving the chance that the
recipient will receive the message as soon as possible.

Users would have the option to exchange information about carriers that they
mutually trust for this purpose.  Trust will still be needed even whilst the
messages are encrypted, not only because carriers might fail to forward the
messages, but also because carriers would be able to observe the presence of a
conversation between sender and recipient, even if the conversation itself is
encrypted.  For this reason, although it is possible for anyone to carry
messages for anyone else, and perhaps even possible for users to act as
``professional carriers'' in exchange for a fee, it is not assumed that senders
would trust arbitrary users to act as carriers, nor is it assumed that
arbitrary users would want to act as carriers for senders (or recipients) that
they do not trust.  We assume that users will keep track of OpenPGP keys and lists
of trusted carriers in local \textit{address books}, as shown in
Figure~\ref{f:book}.  In particular, one particular user might keep track of
the specific user responsible for the introduction to a third user, and the set
of introductions to the same user can form the basis of a list of designated
carriers.

Although we imagine that in most cases a sender would only use a carrier after
failing to deliver the message directly, we assume that a user can opt to send
messages via multiple carriers at the same time, and perhaps also attempt
direct contact with the recipient similarly.  We further assume that the
recipient's software will delete (and possibly track) any duplicate messages.
If a recipient, Carol, knows that a specific third-party, Bob, often acts as a
carrier, then Bob's software and Carol's software may allow Carol to explicitly
ask Bob to forward any messages that he holds for her, although this is not
required.  We assume that carriers will also have the option to inform senders
that a message was sent successfully, although end-to-end delivery confirmation
can be handled by established SMTP headers~\cite{confirmation}.

\subsection{Accounts and Forwarding}

\begin{figure}
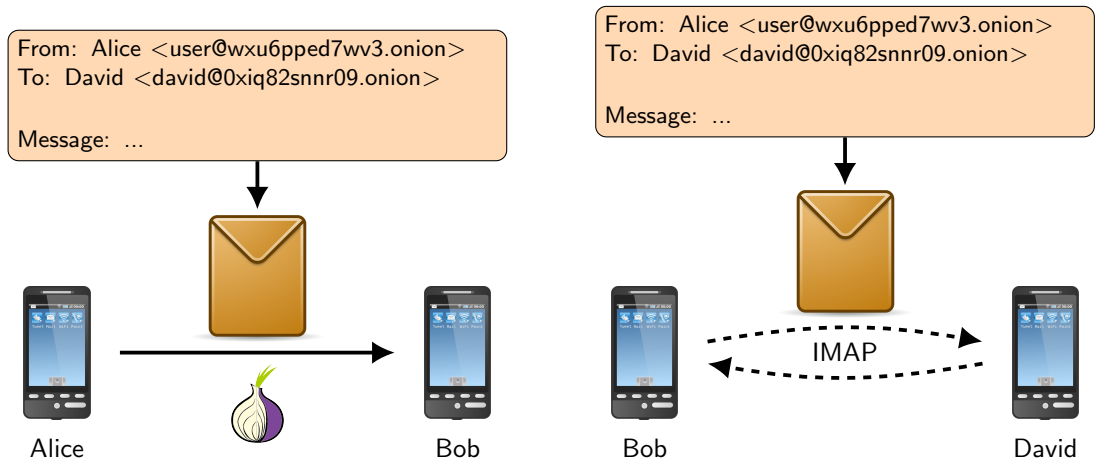

\begin{center}
\begin{tikzpicture}[
    >=latex,
    font=\sf,
    auto
]\ts
\def\smbwd{8em}

\node (r1) at (0,0) [noshape, text width=4em] {
    \scalebox{0.1}{\includegraphics{images/1281043443.png}}
};
\node (r2) at (5,0) [noshape, text width=4em] {
    \scalebox{0.1}{\includegraphics{images/1281043443.png}}
};

\draw[->, line width=0.5mm] (r1) -- node[above] {
    \scalebox{0.1}{\includegraphics{images/rodentia-icons_package-x-generic.png}}
} node[below] {
    \scalebox{0.2}{\includegraphics{images/onion.png}}
} (r2);

\node (m1) at (2.5, 3.4) [
    box,
    yshift=-5pt,
    color=black,
    fill=orange!30
] {
    \small{From: Alice \textless{user@wxu6pped7wv3.onion}\textgreater}\\
    \small{To: David \textless{david@0xiq82snnr09.onion}\textgreater}\\
    \vspace{1.25em}
    \small{Message: ...}
};
\draw[->, line width=0.5mm] (m1) -- (2.5, 1.8);

\node (d1) at (0,-1.2) [nos] {Alice};
\node (d2) at (5,-1.2) [nos] {Bob};

\end{tikzpicture}
\begin{tikzpicture}[
    >=latex,
    font=\sf,
    auto
]\ts
\def\smbwd{8em}

\node (r1) at (0,0) [noshape, text width=4em] {
    \scalebox{0.1}{\includegraphics{images/1281043443.png}}
};
\node (r2) at (5,0) [noshape, text width=4em] {
    \scalebox{0.1}{\includegraphics{images/1281043443.png}}
};

\draw[->, dashed, line width=0.5mm] (r1) edge[bend left=10] node[above] {
    \scalebox{0.1}{\includegraphics{images/rodentia-icons_package-x-generic.png}}
}(r2);

\draw[->, dashed, line width=0.5mm] (r2) edge[bend left=10] node[above, yshift=0.2em] {
    IMAP
}(r1);

\node (m1) at (2.5, 3.7) [
    box,
    yshift=-5pt,
    color=black,
    fill=orange!30
] {
    \small{From: Alice \textless{user@wxu6pped7wv3.onion}\textgreater}\\
    \small{To: David \textless{david@0xiq82snnr09.onion}\textgreater}\\
    \vspace{1.25em}
    \small{Message: ...}
};
\draw[->, line width=0.5mm] (m1) -- (2.5, 2.1);

\node (d1) at (0,-1.2) [nos] {Bob};
\node (d2) at (5,-1.2) [nos] {David};

\end{tikzpicture}

\caption{\textsc{Accounts.} \textit{If desired, Bob can also handle mail for
David directly.  In this setup, Alice sends mail to Bob's device, and David
pulls it from Bob via IMAP (via Tor, or not). }}

\label{f:accounts}
\end{center}
\end{figure}

Users of this messaging system can provide services to communicate with others
who might not run their own mail server software.  For example, a user of the
system can provide an account for someone who might or might not be a user of
the system, as shown in Figure~\ref{f:accounts}.  Mail for the external user
(David, in our example) that is received by the carrier (Bob, in our example)
can be stored in an account on the carrier's device.  Then, the external user
can access the account via an ordinary mail-access protocol such as
IMAP~\cite{imap}, which can be offered to the external user as a Tor onion
service.  (It would also be possible for the carrier to offer the account to
the external user without using a Tor onion service, although this might not
work if the carrier is behind a firewall or NAT.)

Bob might be motivated to provide an account to David for a variety of reasons.
For example, Bob might be David's employer or business partner, or Bob might
offer David an always-on service to receive messages from anonymous senders
more reliably in exchange for a fee.

\begin{figure}
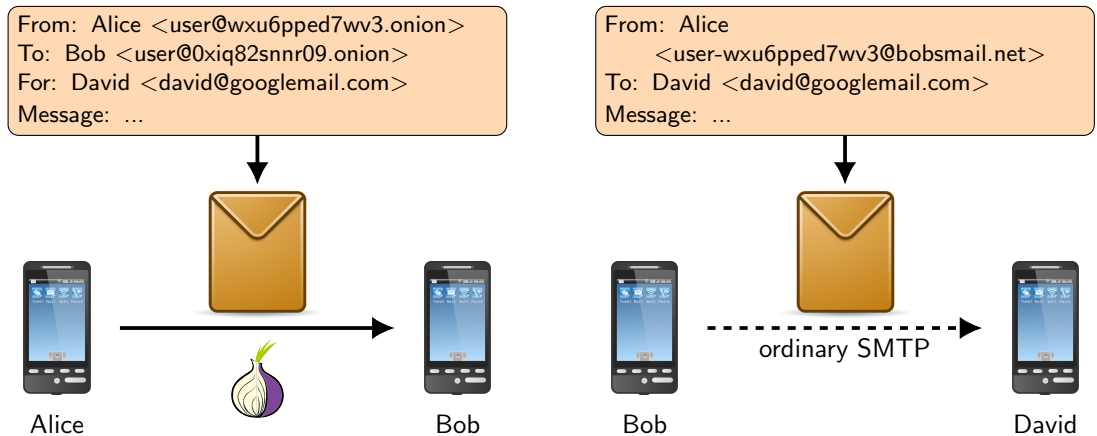

\begin{center}
\begin{tikzpicture}[
    >=latex,
    font=\sf,
    auto
]\ts
\def\smbwd{8em}

\node (r1) at (0,0) [noshape, text width=4em] {
    \scalebox{0.1}{\includegraphics{images/1281043443.png}}
};
\node (r2) at (5,0) [noshape, text width=4em] {
    \scalebox{0.1}{\includegraphics{images/1281043443.png}}
};

\draw[->, line width=0.5mm] (r1) -- node[above] {
    \scalebox{0.1}{\includegraphics{images/rodentia-icons_package-x-generic.png}}
} node[below] {
    \scalebox{0.2}{\includegraphics{images/onion.png}}
} (r2);

\node (m1) at (2.5, 3.4) [
    box,
    yshift=-5pt,
    color=black,
    fill=orange!30
] {
    \small{From: Alice \textless{user@wxu6pped7wv3.onion}\textgreater}\\
    \small{To: Bob \textless{user@0xiq82snnr09.onion}\textgreater}\\
    \small{For: David \textless{david@googlemail.com}\textgreater}\\
    \small{Message: ...}
};
\draw[->, line width=0.5mm] (m1) -- (2.5, 1.8);

\node (d1) at (0,-1.2) [nos] {Alice};
\node (d2) at (5,-1.2) [nos] {Bob};

\end{tikzpicture}
\begin{tikzpicture}[
    >=latex,
    font=\sf,
    auto
]\ts
\def\smbwd{8em}

\node (r1) at (0,0) [noshape, text width=4em] {
    \scalebox{0.1}{\includegraphics{images/1281043443.png}}
};
\node (r2) at (5,0) [noshape, text width=4em] {
    \scalebox{0.1}{\includegraphics{images/1281043443.png}}
};

\draw[->, dashed, line width=0.5mm] (r1) -- node[above] {
    \scalebox{0.1}{\includegraphics{images/rodentia-icons_package-x-generic.png}}
} node[below] {
    ordinary SMTP
} (r2);

\node (m1) at (2.5, 3.4) [
    box,
    yshift=-5pt,
    color=black,
    fill=orange!30
] {
    \small{From: Alice}\\
    \hspace{2em}\small{\textless{user-wxu6pped7wv3@bobsmail.net}\textgreater}\\
    \small{To: David \textless{david@googlemail.com}\textgreater}\\
    \small{Message: ...}
};
\draw[->, line width=0.5mm] (m1) -- (2.5, 1.8);

\node (d1) at (0,-1.2) [nos] {Bob};
\node (d2) at (5,-1.2) [nos] {David};

\end{tikzpicture}

\caption{\textsc{External Forwarding.} \textit{If desired, Bob can deliver mail
to David at an external address.  In this setup, Alice sends mail to Bob's
device, and Bob forwards it to David via ordinary SMTP.  Later, when David
wants to reply to Alice, he is provided an address for Alice that instructs
Bob's mail server to forward it appropriately.}}

\label{f:forwarding}
\end{center}
\end{figure}

Users of this messaging system can also exchange messages with others who
have external e-mail accounts, as shown in Figure~\ref{f:forwarding}.  In
this case, messages for external users (David, in the example) would be
sent to the carrier (Bob, in the example) with a special tag
specifying the external user to which the message must be forwarded.
Then, the SMTP server running on the carrier's device will create a
new message containing the contents of the original message and send it
via ordinary SMTP to the external user.  To facilitate replies, the new
message will provide an address for the sender consisting of a username
encoding the in-system e-mail address of the sender and a hostname
corresponding to the externally reachable address of the carrier's
mail server.

\subsection{Mailing Lists}
\label{ss:lists}

Finally, the system can be used to implement decentralised mailing lists that
do not rely upon a mailing list server.  Such mailing lists would rely upon
pairwise relationships among list members.  We assume that each participant in
a mailing list will have exchanged OpenPGP keys and agreed to receive messages
from at least one of the other participants.  We further assume that the graph
formed by those pairwise connections is fully connected.  Then, a gossip
protocol can be used to share messages through the entire network and, as long
as all participants remain connected and continue to share messages, every
message will eventually reach every user.  Gossip protocols can be optimised to
reduce the total number of messages in several ways, such as using randomised
delays to reduce the chance that messages will cross paths, establishing
agreement among participants to maintain sparseness by policy, or explicitly
selecting a spanning tree~\cite{perlman-sta}.

\begin{figure}
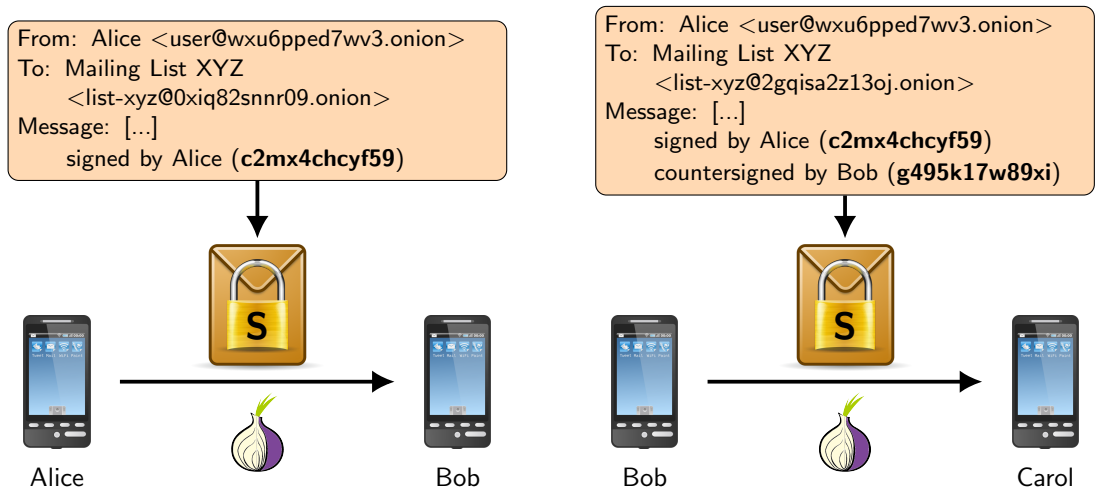

\begin{center}
\begin{tikzpicture}[
    >=latex,
    font=\sf,
    auto
]\ts
\def\smbwd{8em}

\node (r1) at (0,0) [noshape, text width=4em] {
    \scalebox{0.1}{\includegraphics{images/1281043443.png}}
};
\node (r2) at (5,0) [noshape, text width=4em] {
    \scalebox{0.1}{\includegraphics{images/1281043443.png}}
};

\draw[->, line width=0.5mm] (r1) -- node[above] {
    \scalebox{0.1}{\includegraphics{images/rodentia-icons_package-x-generic.png}}
} node[below] {
    \scalebox{0.2}{\includegraphics{images/onion.png}}
} (r2);
\node (i1) at (2.5, 0.95) {
    \scalebox{0.07}{\includegraphics{images/Padlock-gold.png}}
};

\node (m1) at (2.5, 3.7) [
    box,
    yshift=-5pt,
    color=black,
    fill=orange!30
] {
    \small{From: Alice \textless{user@wxu6pped7wv3.onion}\textgreater}\\
    \small{To: Mailing List XYZ}\\
    \hspace{2em}\small{\textless{list-xyz@0xiq82snnr09.onion}\textgreater}\\
    \small{Message: [...]}\\
    \hspace{2em}\small{signed by Alice (\textbf{c2mx4chcyf59})}
};
\draw[->, line width=0.5mm] (m1) -- (2.5, 1.8);

\node (d1) at (0,-1.2) [nos] {Alice};
\node (d2) at (5,-1.2) [nos] {Bob};
\node (d3) at (2.5,0.7) [nos] {\Large{\textbf{S}}};

\end{tikzpicture}
\begin{tikzpicture}[
    >=latex,
    font=\sf,
    auto
]\ts
\def\smbwd{8em}

\node (r1) at (0,0) [noshape, text width=4em] {
    \scalebox{0.1}{\includegraphics{images/1281043443.png}}
};
\node (r2) at (5,0) [noshape, text width=4em] {
    \scalebox{0.1}{\includegraphics{images/1281043443.png}}
};

\draw[->, line width=0.5mm] (r1) -- node[above] {
    \scalebox{0.1}{\includegraphics{images/rodentia-icons_package-x-generic.png}}
} node[below] {
    \scalebox{0.2}{\includegraphics{images/onion.png}}
} (r2);
\node (i1) at (2.5, 0.95) {
    \scalebox{0.07}{\includegraphics{images/Padlock-gold.png}}
};

\node (m1) at (2.5, 3.7) [
    box,
    yshift=-5pt,
    color=black,
    fill=orange!30
] {
    \small{From: Alice \textless{user@wxu6pped7wv3.onion}\textgreater}\\
    \small{To: Mailing List XYZ}\\
    \hspace{2em}\small{\textless{list-xyz@2gqisa2z13oj.onion}\textgreater}\\
    \small{Message: [...]}\\
    \hspace{2em}\small{signed by Alice (\textbf{c2mx4chcyf59})}\\
    \hspace{2em}\small{countersigned by Bob (\textbf{g495k17w89xi})}
};
\draw[->, line width=0.5mm] (m1) -- (2.5, 1.8);

\node (d1) at (0,-1.2) [nos] {Bob};
\node (d2) at (5,-1.2) [nos] {Carol};
\node (d3) at (2.5,0.7) [nos] {\Large{\textbf{S}}};

\end{tikzpicture}

\caption{\textsc{Mailing Lists} \textit{Mailing lists can be implemented using
a gossip protocol among a set of pairwise-linked network list participants.
Members shall forward messages to each other and break cycles by rejecting
duplicates.  Rate-limiting and filtering rules can be implemented by individual
list members using the signature on the message.  Members can also authenticate
message carriers by requiring each message sent to a list to be countersigned
by the peer that forwarded the message.  It is possible to use a distributed
ledger to establish consensus about the set of messages that have been sent to
the list, the sequence of the messages, or the agreed-upon state of a system
based upon the aggregation of the messages. (Note: the padlock with the `S'
symbol indicates that a message is signed.)}}

\label{f:lists}
\end{center}
\end{figure}

Figure~\ref{f:lists} illustrates how a message would be sent and propagated
among a set of list participants.  List members would propagate a message
through the network link by link, each member receiving messages from one of
its neighbours and passing it to its other neighbours, discarding any
duplicates.  We assume that a message received by a list member would be signed
by the original author and that the entire message and the signature by its
original author would be countersigned by the neighbour from which it was
directly received.  Then, the list member can decide whether to propagate the
message based upon its policy and the validity of the signatures.

It is possible to use a distributed ledger to help ensure the completeness of
the list conversation.  For example, a list could require that each message
includes a hash of the previous message, and that an invalid hash would cause a
message to be rejected for propagation by policy.  It is also possible to use a
distributed ledger to keep track of the consensus state of something on behalf
of the entire group.  For example, a distributed ledger could be used to track
and serialise the changes to a document, a record of transactions of tokens, or
a series of calculations.

\section{Security Considerations}

The following is a partial list of security considerations that are important
to the design of systems that make use of the protocol described in this
document:

\begin{enumerate}

\item \textit{Identity revocation.} In the absence of a centralised authority
or revocation server, users will need to share revocation announcements with
their peers.  Two forms of revocation are important: the revocation of a user's
onion service address, and the revocation of a user's OpenPGP key.  Both could
be handled in a peer-to-peer manner by sending revocation messages to a user's
known peers, or through the use of a third-party keyserver.

\item \textit{Managing multiple devices.}  Users could copy their Tor onion
service keys and their OpenPGP keys to multiple devices, although it is the
responsibility of the user to ensure that only one Tor onion service is active
at a given time for a given Tor onion service key.  Users could also use
different keys for different devices and use them as carriers for each other.

\item \textit{Attacks to link multiple identities for a user.}  Users should be
careful to control the set of parties with whom they share their addresses, in
case they want to establish different identities and ensure that they stay
unlinked.  Also, adversaries could employ timing attacks to statistically link
different identities that simultaneously operate on the same device.  The fact
that users can make introductions does not prevent them from having multiple
identities for use with different groups or in different contexts.

\end{enumerate}

\section{Contexts}

Identity, like motion, is not absolute and is always relative.  Put another
way, identity is a matter of perspective.  A simple way to think of the meaning
of \textit{context} is as a `path of connection'.  If Bob introduces Carol to
Alice, then we might say that Alice knows Carol through Bob.  Now, Alice might
already happen to know Carol, or not, but there is no reason why Carol would
use the same identity every time she communicates.  Carol has an incentive to
represent herself differently to everyone that she meets, so that she can
control the linkages among the identities that she uses in different
contexts~\cite{goodell2019}.  Of course, this does not work so well if Bob is
doing all of Carol's introductions for her.  Even if Bob and Carol agree to use
a different identity for Carol in each of the introductions he makes for her,
Bob cannot prove that the Carol he introduces is anyone other than his own
fabrication, and he cannot prove that any two introductions he makes are or are
not the same fabrication.

For this reason, there is value for being introduced to the same person via
multiple channels.  If David and Bob both introduce the same Carol to Alice,
then Alice knows that Carol is someone whose identity David and Bob agree upon.
It could still be that David and Bob have conspired to invent Carol, but that
possibility is less simple than something that they could each have invented
independently.  Conversely, Carol might have one identity that Bob uses to
introduce her to Alice and a different identity that David uses to introduce
her to Alice, in which case Alice might assume that these are two different
people when in fact they are not.

In human interaction, we are generally able to have many different identities.
When I enter a coffee shop, the barista generally does not know any of my other
identities or the link between me and those identities, so I am free to create
an entirely new one.  This is actually a great privilege, since I am unburdened
by other contexts.  Interestingly, it is a privilege that celebrities lack.  If
Boris Johnson were to enter a coffee shop in central London and attempt to
create a new identity, he would be out of luck, precisely because of the
pre-existing context that already binds him.

\section{Acknowledgements}

The author would like to thank Patric de Gentile-Williams for providing
perspective and insight.

\sf

\noindent The Tor onion logo is a registered trademark of the Tor Project, Inc.
All other icons and clipart images are in the public domain.


\begin{thebibliography}{1}\raggedright
\footnotesize

\bibitem{smtp}{
    P. Resnick, Ed.
    ``Internet Message Format.''
    Internet Engineering Task Force RFC 5322,
    October 2008.
    [online]
    \url{https://tools.ietf.org/html/rfc5322}
    [retrieved 2020-07-03]
}
\bibitem{tor}{
    R. Dingledine, N. Mathewson, and P. Syverson.
    ``Tor: The Second-Generation Onion Router.''
    \textit{Proceedings of the 13th USENIX Security Symposium},
    2004.
    [online]
    \url{https://www.nrl.navy.mil/itd/chacs/sites/www.nrl.navy.mil.itd.chacs/files/pdfs/Dingledine%20etal2004.pdf}
    [retrieved 2018-10-10]
}
\bibitem{imap}{
    M. Crispin.
    ``Internet Message Access Protocol.''
    Internet Engineering Task Force RFC 3501,
    March 2003.
    [online]
    \url{https://tools.ietf.org/html/rfc3501}
    [retrieved 2020-07-03]
}
\bibitem{tor-onion}{
    D. Goulet, G. Kadianakis, and N. Mathewson.
    ``Next-Generation Hidden Services in Tor.''
    2013-11-29.
    [online]
    \url{https://gitweb.torproject.org/torspec.git/tree/proposals/224-rend-spec-ng.txt}
    [retrieved 2020-07-03]
}
\bibitem{spf}{
    S. Kitterman.
    ``Sender Policy Framework (SPF) for Authorizing Use of Domains in Email, Version 1.''
    Internet Engineering Task Force RFC 7208,
    April 2014.
    [online]
    \url{https://tools.ietf.org/html/rfc7208}
    [retrieved 2020-07-03]
}
\bibitem{dkim}{
    D. Crocker, T. Hansen, and M. Kucherawy, Eds.
    ``DomainKeys Identified Mail (DKIM) Signatures.''
    Internet Engineering Task Force RFC 6376,
    September 2011.
    [online]
    \url{https://tools.ietf.org/html/rfc6376}
    [retrieved 2020-07-03]
}
\bibitem{mixminion}{
    G. Danezis, R. Dingledine, and N. Mathewson.
    ``Mixminion: Design of a Type III Anonymous Remailer Protocol.''
    \textit{Proceedings of the 2003 IEEE Symposium on Security and Privacy},
    May 2003.
    [online]
    \url{https://www.mixminion.net/minion-design.pdf}
    [retrieved 2020-07-03]
}
\bibitem{mixmaster}{
    U. M\"oller, L Cottrell, P. Palfrader, and L. Sassaman.
    ``Mixmaster Protocol Version 2.''
    Internet Engineering Task Force Internet-Draft,
    2004-12-29.
    [online]
    \url{https://tools.ietf.org/html/draft-sassaman-mixmaster-03}
    [retrieved 2020-07-03]
}
\bibitem{mixminion-web}{
    N. Mathewson.
    ``Mixminion: A Type III Anonymous Remailer.''
    [online]
    \url{https://www.mixminion.net/}
    [retrieved 2020-07-03]
}
\bibitem{cwtch}{
    S. Lewis.
    ``Cwtch: Privacy Preserving Infrastructure for Asynchronous, Decentralized, Multi-Party and Metadata Resistant Applications.''
    Discussion Paper,
    2018-06-28.
    [online]
    \url{https://cwtch.im/cwtch.pdf}
    [retrieved 2020-07-03]
}
\bibitem{ricochet}{
    R. Burchell.
    Ricochet Protocol.
    [online]
    \url{https://github.com/ricochet-im/ricochet/blob/master/doc/protocol.md}
    [retrieved 2020-07-03]
}
\bibitem{greene2017}{
    D. Greene and K. Shilton.
    ``Platform privacies: Governance, collaboration, and the different meanings of `privacy' in iOS and Android development.''
    \textit{New Media \& Society} \textbf{20}(4),
    pp. 1640--1657,
    2017-04-27.
    [online]
    \url{https://journals.sagepub.com/doi/pdf/10.1177/1461444817702397}
    [retrieved 2020-07-13]
}
\bibitem{tor-control}{
    Tor Project, Inc.
    ``TC: A Tor control protocol (Version 1).''
    [online]
    \url{https://gitweb.torproject.org/torspec.git/tree/control-spec.txt}
    [retrieved 2020-07-03]
}
\bibitem{zimmermann1991}{
    P. Zimmermann.
    ``Why I Wrote PGP.''
    \textit{PGP User's Guide},
    1991.
    [online]
    \url{https://www.philzimmermann.com/EN/essays/WhyIWrotePGP.html}
    [retrieved 2018-10-11]
}
\bibitem{gpg}{
    J. Callas, L. Donnerhacke, H. Finney, D. Shaw, and R. Thayer.
    ``OpenPGP Message Format.''
    Internet Engineering Task Force RFC 4880,
    November 2007.
    [online]
    \url{https://tools.ietf.org/html/rfc4880}
    [retrieved 2020-07-03]
}
\bibitem{confirmation}{
    K. Moore.
    ``Simple Mail Transfer Protocol (SMTP) Service Extension for Delivery Status Notifications (DSNs).''
    Internet Engineering Task Force RFC 3461,
    January 2003.
    [online]
    \url{https://tools.ietf.org/html/rfc3461}
    [retrieved 2020-07-03]
}
\bibitem{perlman-sta}{
    R. Perlman.
    ``An algorithm for distributed computation of a spanningtree in an extended LAN.''
    ACM SIGCOMM Computer Communication Review,
    September 1985.
    \url{doi:10.1145/319056.319004}
    [online]
    \url{https://www.it.uu.se/edu/course/homepage/datakom/ht06/slides/sta-perlman.pdf}
    [retreived 2020-07-13]
}
\bibitem{goodell2019}{
    G. Goodell and T. Aste.
    ``A Decentralised Digital Identity Architecture.''
    \textit{Frontiers in Blockchain},
    2019-11-05.
    \url{doi:10.3389/fbloc.2019.00017}.  Also available on arXiv:
    \url{https://arxiv.org/pdf/1902.08769}
}

\end{thebibliography}
\end{document}